# The importance of scale in spatially varying coefficient modeling

Daisuke Murakami (Corresponding author)
Department of Statistical Modeling, The Institute of Statistical Mathematics,
10-3 Midori-cho, Tachikawa, Tokyo 190-8562, Japan
E-mail: dmuraka@ism.ac.jp

Binbin Lu
School of Remote Sensing and Information Engineering, Wuhan University,
Wuhan, Hubei, China

Paul Harris
Sustainable Agricultural Sciences, Rothamsted Research,
North Wyke, Okehampton, UK

Chris Brunsdon
National Centre for Geocomputation, Maynooth University,
Maynooth, Kildare, Ireland

Martin Charlton
National Centre for Geocomputation, Maynooth University,
Maynooth, Kildare, Ireland

Tomoki Nakaya
Department of Geography, Ritsumeikan University,
Kyoto, Kyoto, Japan

Daniel A. Griffith
School of Economic, Political and Policy Sciences, The University of Texas at Dallas,
Dallas, Texas, USA

**Abstract**: While spatially varying coefficient (SVC) models have attracted considerable attention in applied science, they have been criticized as being unstable. The objective of this study is to show that capturing the "spatial scale" of each data relationship is crucially important to make SVC modeling more stable, and in doing so, adds flexibility. Here, the analytical properties of six SVC models are summarized in terms of their characterization of scale. Models are examined through a series of Monte Carlo simulation experiments to assess the extent to which spatial scale influences model stability and the accuracy of their SVC estimates. The following models are studied: (i) geographically weighted regression (GWR) with a fixed distance or (ii) an adaptive distance bandwidth (GWRa), (iii) flexible bandwidth GWR (FB-GWR) with fixed distance or (iv) adaptive distance bandwidths (FB-GWRa), (v) eigenvector spatial filtering (ESF), and (vi) random effects ESF (RE-ESF). Results reveal that the SVC models designed to capture scale dependencies in local relationships (FB-GWR, FB-GWRa and RE-ESF) most accurately estimate the simulated SVCs, where RE-ESF is the most computationally efficient. Conversely GWR and ESF, where SVC estimates are naively assumed to operate at the same spatial scale for each relationship, perform poorly. Results also confirm that the adaptive bandwidth GWR models (GWRa and FB-GWRa) are superior to their fixed bandwidth counterparts (GWR and FB-GWR).

1. Introduction

Spatially varying coefficient (SVC) models are used to investigate non-stationarity in response to predictor data relationships in regression models. Provided relationship heterogeneity exists, models will output regression coefficients that vary across space. These SVCs can be mapped along with associated inference diagnostics, and thus provide a deeper understanding of a study's spatial relationships. As with spatial autocorrelation, relationship spatial heterogeneity is a common property of many geographical processes (see Anselin, 2010), although differentiating one effect from the other can be difficult (e.g., Harris et al., 2017). Various approaches have been developed for SVC regression modelling, the most notable of which include: (i) the spatial expansion method (Casetti, 1972; Casetti and Jones, 1992), (ii) geographically weighted regression (GWR, Brunsdon et al., 1996; 1998; Fotheringham et al., 2002), (iii) Bayesian SVC models (Gelfand, et al., 2003; Gamerman et al., 2003; Assunçao, 2003; Wheeler and Calder, 2007; Wheeler and Waller, 2009), and (iv) eigenvector spatial filtering (ESF)- based approaches (Griffith, 2003; 2008; Murakami et al., 2017).

Among them, GWR has proven the most popular, including case studies in hedonic house price modelling (e.g., Bitter et al., 2007; Páez et al., 2008; Lu et al., 2014a), environmental analysis (e.g., Brunsdon et al., 2001; Jaimes et al., 2010; Harris et al., 2010a) and disease mapping (e.g., Nakaya et al., 2005; Hu et al. 2012; Ndiath et al., 2015). Much of this popularity stems from its relative simplicity and readily-available software like GWR4 (Nakaya, 2015, http://gwr.maynoothuniversity.ie/gwr4-software/) and the **GWmodel** R package (Lu et al., 2014b; Gollini et al., 2015)). Despite the wide-spread uptake of basic GWR, it suffers from (at least) two severe limitations: (a) instability where local predictor variable collinearity can create spurious non-stationarities (Wheeler and Tiefelsdorf, 2005; Paez et al., 2011), and (b) inflexibility where basic GWR assumes the same scale of spatial variation across each set of estimated SVCs (Brunsdon et al., 1999).

Relating model limitation (a), it has been demonstrated that SVCs estimated from GWR can be collinear each other, detect unrealistically smooth map patterns, and/or take extreme values. Various collinearity diagnostics can be calculated to provide a better understanding of potential problems (Wheeler and Tiefelsdorf, 2005; Wheeler 2007; Gollini et al., 2015), together with the implementation of some regularized GWR model (Wheeler, 2007; 2009; Gollini et al., 2015) or GWR via an empirical Bayes approach (Bárcena et al., 2014) – all specifically designed to address collinearity. It has been argued that GWR is actually fairly robust to local collinear effects (Fotheringham and Oshan, 2016), but on balance, evidence suggests otherwise (e.g., see Harris et al., 2017). Observe that instability in estimated SVCs from GWR may arise for other reasons than that due to collinearity, including the existence of outliers (Farber and Páez, 2007; Harris et al., 2010a) and those due to spatial autocorrelation (Cho et al., 2010).

For model limitation (b), basic GWR uses a single kernel bandwidth for its calibration, which is somewhat flawed in that it implicitly assumes the same degree of spatial smoothness for each set of SVCs, which is unrealistic. Thus, when some relationships tend to operate at a larger-scale whilst other relationships operate at a smaller-scale, basic GWR will nullify these differences and only find a 'best-on-average' scale of relationship non-stationarity (as using only a single bandwidth). To address this limitation, mixed (semiparametric) GWR can be implemented in which some relationships are assumed stationary (globally-fixed) whilst others are assumed non-stationary (locally-varying) (Brunsdon et al., 1999; Fotheringham et al., 2002; Nakaya et al., 2005; Mei et al., 2006; 2016). However, a mixed GWR model only in part addresses the limitation, as the subset of locally-varying relationships is still assumed to operate at the same spatial scale. Instead flexible bandwidth GWR (FB-GWR) can be used, in which each relationship is specified using its own bandwidth, and thus provides a true multi-scale GWR model, where the scale of relationship non-stationarity may vary for each response to predictor variable relationship.

The development of FB-GWR follows that of Yang et al. (2011; 2012); Yang (2014); Lu et al. (2017); Leong and Yue (2017) (who re-name it conditional GWR) and Fotheringham et al. (2017) (who re-name it multiscale GWR), all of whom implement the idea of 'a vector of bandwidths' for GWR, as first set out in Brunsdon et al. (1999). The study of Lu et al. (2017) provides an extension of FB-GWR, where each relationship can also be specified with its own distance metric, as well as its own bandwidth. In this study, we implement the model of Lu et al. (2017), but where it is specified using only Euclidean distances – thus directly providing a FB-GWR model.

A Bayesian SVC model (specifically, the geostatistical approach of Gelfand et al., 2003) can be viewed as a regularized alternative to GWR (i.e., it can address collinearity), and is also directly able to identify the spatial scale of each relationship through its specification of geostatistical priors. Thus, model limitations (a) and (b) stated for GWR are implicitly addressed. However, although an increased coefficient accuracy for the Bayesian SVC approach has been reported (see Wheeler and Calder, 2007; Wheeler and Waller, 2009), it is computationally expensive, especially when scale is estimated for every set of SVCs (Finley, 2011).

Unlike GWR, the ESF-based approach allows for controlling the number of parameters (i.e., model complexity) through variable selection. However, Helbich and Griffith (2016); Murakami et al. (2017); Oshan and Fotheringham (2017) all demonstrate the instability of the ESF-based approach where it can suffer just as basic GWR does with respect to limitations (a) and (b). In this respect, Murakami et al. (2017) propose an extended ESF-based approach that directly addresses limitations (a) and (b) (i.e., it is robust to local collinearity and allows the possibility for each set of SVCs to have a different degree of spatial smoothness). Furthermore, this random effects ESF model (RE-ESF) is shown to be computationally efficient, thus providing a real-world alternative to the Bayesian SVC model that is often computationally intractable.

The study of Murakami et al. (2017), through a Monte Carlo simulation experiment similar in design to that used here, not only demonstrated the advantage of the RE-ESF model over the Bayesian SVC model, but also demonstrated its advantages over both basic and regularized GWR forms (following Gollini et al., 2015). The latter was not surprising given that neither single-bandwidth GWR forms address model limitation (b). Hence, this study addresses this important gap by introducing FB-GWR to the same model comparison exercise. Because it is unnecessary to repeat all model comparisons of Murakami et al. (2017), only (1) GWR with a fixed distance or (2) an adaptive distance bandwidth (GWRa), (3) FB-GWR with fixed distance or (4) adaptive distance bandwidths (FB-GWRa), (5) ESF, and (6) RE-ESF models are compared here. Thus, this study taken together with that of Murakami et al. (2017) provides a comprehensive comparison of all known multi-scale SVC models (i.e., FB-GWR, RE-ESF and Bayesian SVC models).

In summary, the aim of this study is to continue to demonstrate the importance of "spatial scale" in SVC models through FB-GWR and RE-ESF, whose outputs should be more stable and flexible in comparison to their basic counterparts (GWR and ESF, respectively). The remaining sections are organized as follows. Section 2 outlines the GWR- and ESF-based models; section 3 performs a Monte Carlo simulation experiment to quantify the impact of spatial scale on model stability; section 4 summarizes a second Monte Carlo simulation experiment to evaluate the impact of spatial scale on SVC estimates; and section 5 provides a concluding discussion. Study GWR and FB-GWR models are fitted using **GWmodel** (version 2.0.4.; https://cran.r-project.org/package=GWmodel), the RE-ESF model is fitted using the R package **spmoran** (version 0.1.2.; Murakami, 2017; https://cran.r-project.org/web/packages/spmoran/index.html), whilst the ESF model is fitted by newly written R codes.

## 2. Spatially varying coefficient modeling

### 2.1. The over-arching SVC model

A linear SVC model is formulated as follows:

$$y_i = \sum_{k=1}^{K} x_{i,k}\beta_{i,k} + \varepsilon_i, \qquad E[\varepsilon_i] = 0, \qquad Var[\varepsilon_i] = \sigma^2, \tag{1}$$

where $y_i$ represents the response variable at the $i$-th sample site, where $i \in \{1 \cdots N\}$, $x_{i,k}$ represents the $k$-th predictor variable, with $k \in \{1 \cdots K\}$, $\varepsilon_i$ represents the disturbance, and $\sigma^2$ represents a variance parameter. $\beta_k(s_i)$ denotes the $k$-th SVC for site $i$. There are *local* and *global* approaches to estimate Eq. (1), as detailed in sections 2.2 and 2.3, where, in general, a *global* approach to non-stationary modelling is preferred as it is more statistically-coherent (e.g., Sampson et al., 2001).

### 2.2. Local estimation (GWR and FB-GWR)

A *local* approach estimates coefficients at the $i$-th site, $\{\beta_1(s_i),... \beta_k(s_i),... \beta_K(s_i)\}$, using only neighboring sub-samples. Moving window regression (MWR; see Lloyd, 2010) applies ordinary least squares estimation to neighboring sub-samples at site $i$, whereas GWR applies weighted least squares estimation to neighboring sub-samples that are weighted via a distance-decay scheme at site *i*. MWR is a special case of GWR when a box-car kernel weighting scheme is specified (weights equal unity within the kernel and zero otherwise). Distance-decay weighting provides added flexibility to local regression modelling, allowing more data to have an influence locally, and tends to yield more smoothly-varying coefficient surfaces. Suppose that $\boldsymbol{\beta}(s_i) = [\beta_1(s_i),... \beta_k(s_i),... \beta_K(s_i)]'$, where " ′ " represents matrix transpose, the GWR estimator yields:

$$\hat{\boldsymbol{\beta}}(s_i) = [\mathbf{X}'\mathbf{G}(s_i)\mathbf{X}]^{-1}\mathbf{X}'\mathbf{G}(s_i)\mathbf{y} \qquad (2)$$

where **X** is an $N \times K$ matrix of predictor variables, **y** is an $N \times 1$ vector of continuous response variables, and $\mathbf{G}(s_i)$ is an $N \times N$ diagonal matrix whose *j*-th element $g(s_i, s_j)$ represents the weight assigned to the *j*-th sample. Here, $g(s_i, s_j)$ is calculated by some kernel weighting function (see Gollini et al., 2015). For instance, the exponential kernel is defined as follows:

$$g(s_i, s_j) = \exp\left(-\frac{d(s_i, s_j)}{b}\right) \qquad (3)$$

where $d(s_i, s_j)$ is the distance between locations $s_i$ and $s_j$, and b denotes the bandwidth parameter. The resultant SVCs tend to the global coefficients of a standard regression, if the bandwidth parameter, *b*, is set sufficiently large enough; otherwise, the SVCs are local. Here the bandwidth can be specified as a fixed distance, but for irregular sample configurations, the kernel window tends to include too few samples in sparsely sampled areas, and too many samples in densely sampled areas. To counter this, an adaptive distance bandwidth can be specified, where the bandwidth varies according to a fixed local density of sub-samples. An adaptive exponential kernel is defined as follows:

$$g^{ad}(s_i, s_j) = \exp\left(-\frac{d(s_i, s_j)}{b^{ad}}\right) \tag{4}$$

where $b^{ad}$ is the adaptive bandwidth for the $i$-th site, and is given by the distance between the $i$-th site and the $j$-th nearest neighbor.

Standard GWR as described above, ignores differences of spatial scale across the SVCs, as the same (single - fixed or adaptive) bandwidth is specified for all data relationships. To counter this, each set of SVCs can be found using its own bandwidth, and thus provide an extension of GWR with multiple bandwidths, one for each relationship (i.e., FB-GWR). Here the fixed bandwidth, exponential kernel for FB-GWR is defined as:

$$g_k(s_i, s_j) = \exp\left(-\frac{d(s_i, s_j)}{b_k}\right), \tag{5}$$

where $b_k$ is the fixed bandwidth for the $k$-th parameter. The $k$-th coefficient estimates may have global scale spatial variations if $b_k$ is set sufficiently large, and local scale spatial variations if $b_k$ is set sufficiently small. The corresponding adaptive bandwidth version for FB-GWR (i.e., for FB-GWRa) is defined as:

$$g_k^{ad}(s_i, s_j) = \exp\left(-\frac{d(s_i, s_j)}{b_k^{ad}}\right), \qquad (6)$$

where $b_k^{ad}$ is the $k$-th adaptive bandwidth.

Standard GWR is estimated as follows: (i) the bandwidth parameter is calibrated by minimizing the mean squared error (MSE) by applying a leave-one-out cross-validation (CV) procedure (Brunsdon et al., 1996); (ii) the SVCs are estimated by substituting the calibrated bandwidth into Eq. (2). FB-GWR is estimated in a similar fashion except for step (i), in which a back-fitting approach is adopted (for details see Lu et al., 2017), which sequentially iterates the calibration of $b_k$ (or $b_k^{ad}$) assuming that all bandwidth parameters are known (see also, Yang, 2014). The MSE minimization in step (i) for GWR or FB-GWR can be replaced with the maximization of the corrected Akaike Information Criterion (AICc), or some other information criterion. Observe that as $N \times K$ coefficients are estimated using only $N$ samples, it is necessary to enhance model accuracy while avoiding over-fitting. A CV or AICc approach is reasonable because it minimizes the generalization error (see Bishop 2006). In this study, the AICc approach is chosen for all GWR and FB-GWR fits, and as detailed above, only bandwidths corresponding to exponential kernels are found.

### 2.3. Global estimation (ESF and RE-ESF)

This *global* approach estimates the SVCs by fitting spatial process models. The spatial expansion and ESF-based approaches are representative of such methods, where the former fits trend surface models, whereas the latter fits ESF models describing spatially structured SVC map patterns. The ESF-based approach is built on the Moran coefficient (MC; see, Cliff and Ord 1973),[1] which is a diagnostic statistic for spatial dependence. The MC is formulated as follows:

$$MC[\mathbf{y}] = \frac{N}{\mathbf{1}'\mathbf{C1}} \frac{\mathbf{y}'\mathbf{MCMy}}{\mathbf{y}'\mathbf{My}}, \tag{7}$$

where $\mathbf{1}$ is an $N \times 1$ vector of ones, $\mathbf{C}$ is an $N \times N$ connectivity matrix whose diagonal elements are zero, and $\mathbf{M} = \mathbf{I} - \mathbf{11}'/N$ is an $N \times N$ centering matrix. The MC is greater than $-1/(N-1) \approx 0$, which is the expectation of the MC in absence of spatial dependence, if the samples are positively spatially dependent, and smaller than $-1/(N-1)$ if they are negatively dependent.[2] Let us eigen-decompose the matrix $\mathbf{MCM}$ to $\mathbf{E}_{full}\mathbf{\Lambda}_{full}\mathbf{E}_{full}'$, where $\mathbf{E}_{full}$ is an $N \times N$ matrix with its *l*-th column being the *l*-th eigenvector $\mathbf{e}_l$, and $\mathbf{\Lambda}_{full}$ is an $N \times N$ diagonal matrix whose *l*-th element is the *l*-the eigenvalue, $\lambda_l$. The eigenvectors have the following feature:

$$\begin{aligned} MC[\mathbf{e}_l] &= \frac{N}{\mathbf{1}'\mathbf{C1}} \frac{\mathbf{e}_l'\mathbf{MCMe}_l}{\mathbf{e}_l'\mathbf{Me}_l} = \frac{N}{\mathbf{1}'\mathbf{C1}} \frac{\mathbf{e}_l'\mathbf{E}_{full}\mathbf{\Lambda}_{full}\mathbf{E}'_{full}\mathbf{e}_l}{\mathbf{e}_l'\mathbf{e}_l}, \\ &= \frac{N}{\mathbf{1}'\mathbf{C1}} \lambda_l. \end{aligned} \tag{8}$$

---

[1] Griffith (2017) shows that the MC base is superior to the GR base, which could be used.

[2]

Here Eq. (8) suggests that the eigenvectors corresponding to positive eigenvalues are orthogonal basis functions describing positive spatial dependence, with each magnitude being indexed by its corresponding eigenvalue. Likewise, eigenvectors corresponding to negative eigenvalues explain negative spatial dependence. For details on Moran eigenvectors, see Griffith (2003).

The ESF-based SVC model of Griffith (2008) is formulated as:

$$\mathbf{y} = \sum_{k=1}^{K} \mathbf{x}_k \circ \boldsymbol{\beta}_k^{ESF} + \boldsymbol{\varepsilon}, \qquad \boldsymbol{\varepsilon} \sim N(\mathbf{0}, \sigma^2 \mathbf{I}),$$
$$\boldsymbol{\beta}_k^{ESF} = \beta_k \mathbf{1} + \mathbf{E}_k \boldsymbol{\gamma}_k, \qquad (9)$$

where $\mathbf{x}_k$ is an $N \times 1$ vector of the $k$-th predictor variable (i.e., the $k$-th column of matrix $\mathbf{X}$), $\mathbf{E}_k$ is an $N \times L_k$ matrix composed of $L_k$ eigenvectors ($L_k < N$), $\boldsymbol{\gamma}_k$ is an $L_k \times 1$ coefficient vector, and "$\circ$" denotes the element-wise (Hadamard) product operator. Here $\boldsymbol{\beta}_k^{ESF} = \beta_k \mathbf{1} + \mathbf{E}_k \boldsymbol{\gamma}_k$ yields a vector of SVCs in which $\beta_k \mathbf{1}$ and $\mathbf{E}_k \boldsymbol{\gamma}_k$ represent the constant component and the spatially varying component, respectively.

The parameters of this model are estimated as follows: (a) eigenvectors, which are not of interest, are removed a priori from $\{\mathbf{E}_1, ..., \mathbf{E}_K\}$ (see below); (b) significant predictor variables are selected among $\{\mathbf{X}, \mathbf{x}_1 \circ \mathbf{E}_1, ..., \mathbf{x}_K \circ \mathbf{E}_K\}$ by applying a forward variable selection technique; (c) $\{\beta_1, ..., \beta_K, \boldsymbol{\gamma}_1, ... \boldsymbol{\gamma}_K\}$ are estimated using the model after the variable selection; and, (d) $\widehat{\boldsymbol{\beta}}_k^{ESF} = \hat{\beta}_k \mathbf{1} + \mathbf{E}_k \hat{\boldsymbol{\gamma}}_k$ is calculated. In our analysis, $\mathbf{E}_k$ is defined by the eigenvectors corresponding to positive eigenvalues in step (a) (see Murakami et al., 2017). Thus, all eigenvectors describing positive spatial dependence are taken into account. The adjusted *R*-square is maximized in the variable selection step (c).

The ESF-based approach, which estimates *deterministic* map patterns, has been extended to a random effects ESF-based approach (RE-ESF; Murakami and Griffith, 2015), which models *stochastic* spatial processes. The RE-ESF-based SVC model (Murakami et al., 2017) is formulated as follows:

$$\mathbf{y} = \sum_{k=1}^{K} \mathbf{x}_k \circ \boldsymbol{\beta}_k^{RE-ESF} + \boldsymbol{\varepsilon}, \qquad \boldsymbol{\varepsilon} \sim N(\mathbf{0}, \sigma^2 \mathbf{I}),$$
$$\boldsymbol{\beta}_k^{RE-ESF} = \beta_k \mathbf{1} + \mathbf{E}_k \boldsymbol{\gamma}_k, \qquad \boldsymbol{\gamma}_k \sim N\left(\mathbf{0}_L, \sigma_{\gamma,k}^2 \boldsymbol{\Lambda}(\alpha_k)\right), \qquad (10)$$

where $\mathbf{0}_L$ is an $L \times 1$ vector of zeros, $\mathbf{E}$ is a matrix of $L$ eigenvectors corresponding to positive eigenvalues, $\sigma_{\gamma,k}^2$ is a variance parameter, and $\boldsymbol{\Lambda}(\alpha_k)$ is an $L \times L$ diagonal matrix whose *l*-th element is $\lambda_l(\alpha_k) = \left(\sum_l \lambda_l / \sum_l \lambda_l^{\alpha_k}\right)\lambda_l^{\alpha_k}$, where $\alpha_k$ is the key parameter. When $\alpha_k$ is large, coefficients of the non-principal eigenvectors are strongly shrunk toward 0, and the *k*-th SVCs, $\boldsymbol{\beta}_k^{RE-ESF}$, provide a large-scale spatial pattern. By contrast, $\boldsymbol{\beta}_k^{RE-ESF}$ has a small-scale spatial pattern when $\alpha_k$ is small. Thus, $\alpha_k$ is a scale parameter for the SVCs, and its effects for RE-ESF are analogous to the multiple bandwidths of FB-GWR.

Furthermore, Eq. (10) has the following expression:

$$\mathbf{y} = \mathbf{X}\boldsymbol{\beta} + \tilde{\mathbf{E}}\tilde{\boldsymbol{\Lambda}}(\boldsymbol{\theta})\tilde{\mathbf{u}} + \boldsymbol{\varepsilon}, \qquad \boldsymbol{\varepsilon} \sim N(\mathbf{0}, \sigma^2\mathbf{I}),$$

$$\tilde{\mathbf{E}} = [\mathbf{x}_1 \circ \mathbf{E} \quad \dots \quad \mathbf{x}_K \circ \mathbf{E}], \quad \tilde{\boldsymbol{\Lambda}}(\boldsymbol{\theta}) = \begin{bmatrix} \sigma_{\gamma,1}^2 \boldsymbol{\Lambda}(\alpha_1) & & \\ & \ddots & \\ & & \sigma_{\gamma,K}^2 \boldsymbol{\Lambda}(\alpha_K) \end{bmatrix}, \quad \tilde{\mathbf{u}} = \begin{bmatrix} \mathbf{u}_1 \\ \vdots \\ \mathbf{u}_K \end{bmatrix}. \tag{11}$$

$\boldsymbol{\theta} \in \{\alpha_1, \cdots \alpha_K, \sigma_{\gamma,1}^2, \cdots \sigma_{\gamma,K}^2\}$, and $\mathbf{u}_k \sim N(\mathbf{0}_L, \sigma^2 \mathbf{I}_L)$, where $\mathbf{I}_L$ is a $L \times L$ identity matrix. Note that $\boldsymbol{\gamma}_k = \sigma_{k(\gamma)}^2 \boldsymbol{\Lambda}(\alpha_k)\mathbf{u}_k$, where Eq. (11) suggests that the RE-ESF model is a linear mixed effects model. Furthermore, $\boldsymbol{\beta}$ and $\tilde{\mathbf{u}}$ have the following best linear unbiased estimators:

$$\begin{bmatrix} \hat{\boldsymbol{\beta}} \\ \hat{\tilde{\mathbf{u}}} \end{bmatrix} = \begin{bmatrix} \mathbf{X}'\mathbf{X} & \mathbf{X}'\tilde{\mathbf{E}}\tilde{\boldsymbol{\Lambda}}(\boldsymbol{\theta}) \\ \tilde{\boldsymbol{\Lambda}}(\boldsymbol{\theta})\tilde{\mathbf{E}}'\mathbf{X} & \tilde{\boldsymbol{\Lambda}}(\boldsymbol{\theta})\tilde{\mathbf{E}}'\tilde{\mathbf{E}}\tilde{\boldsymbol{\Lambda}}(\boldsymbol{\theta}) + \mathbf{I}_{KL} \end{bmatrix}^{-1} \begin{bmatrix} \mathbf{X}'\mathbf{y} \\ \tilde{\boldsymbol{\Lambda}}(\boldsymbol{\theta})\tilde{\mathbf{E}}'\mathbf{y} \end{bmatrix}, \tag{12}$$

where $\boldsymbol{\theta}$ is estimated by numerically maximizing the following type II restricted likelihood (empirical Bayes/$h$-likelihood):

$$\begin{aligned} loglik_R(\boldsymbol{\theta}) = &-\frac{1}{2} log \begin{vmatrix} \mathbf{X}'\mathbf{X} & \mathbf{X}'\tilde{\mathbf{E}}\tilde{\boldsymbol{\Lambda}}(\boldsymbol{\theta}) \\ \tilde{\boldsymbol{\Lambda}}(\boldsymbol{\theta})\tilde{\mathbf{E}}'\mathbf{X} & \tilde{\boldsymbol{\Lambda}}(\boldsymbol{\theta})\tilde{\mathbf{E}}'\tilde{\mathbf{E}}\tilde{\boldsymbol{\Lambda}}(\boldsymbol{\theta}) + \mathbf{I}_{KL} \end{vmatrix} \\ &- \frac{N-K}{2}\left[1 + log\left(\frac{2\pi}{N-K}(\hat{\boldsymbol{\varepsilon}}'\hat{\boldsymbol{\varepsilon}} + \hat{\tilde{\mathbf{u}}}'\hat{\tilde{\mathbf{u}}})\right)\right], \end{aligned} \tag{13}$$

where $\hat{\boldsymbol{\varepsilon}} = \mathbf{y} - \mathbf{X}\hat{\boldsymbol{\beta}} - \tilde{\mathbf{E}}\tilde{\boldsymbol{\Lambda}}(\boldsymbol{\theta})\hat{\tilde{\mathbf{u}}}$. Given $\mathbf{X}'\mathbf{X}$, $\mathbf{X}'\tilde{\mathbf{E}}$, and $\tilde{\mathbf{E}}'\tilde{\mathbf{E}}$, the computational complexity of Eq. (13) is $O((K+KL)^3)$, which is independent of $N$. This ensures that, once these matrix products are evaluated a priori, the numerical optimization of $\boldsymbol{\theta}$ is fast, even for large samples.

### 2.4. The effective number of parameters for the SVC models

This section defines the effective number of parameters, $p^*$, for the study SVC

models, which is a measure of model complexity. For a linear model, $p^*$ is defined by $tr[\mathbf{H}]$, where $tr[\cdot]$ is the trace operator and $\mathbf{H}$ is the hat matrix such that $\hat{\mathbf{y}} = \mathbf{Hy}$. For instance, $p^*$ for the standard linear regression model is $p^*_{LM} = tr[\mathbf{H}_{LM}]$, which equals the number of regression coefficients, where $\mathbf{H}_{LM} = \mathbf{X}(\mathbf{X}'\mathbf{X})^{-1}\mathbf{X}'$. Small $p^*$ is desirable to avoid over-fitting.

[Table 1 around here]

Table 1 summarizes $p^*$ for the study SVC models. Here, $p^*_{GWR} = tr[\mathbf{H}_{GWR}]$ inflates when $\mathbf{X}'\mathbf{G}(s_i)\mathbf{X}$ is nearly singular. Singularity happens when the bandwidth is small and most elements of $\mathbf{G}(s_i)$ take near zero values. In other words, small bandwidths introduce over-fitting. The problem is serious if sub-samples are sparsely distributed around the site $s_i$. Thus, GWR specified with an adaptive bandwidth (GWRa), which changes the kernel window size in accordance with sample density, would be an effective tool to mitigate this problem, where $p^*_{GWRa}$ is likely to be smaller than $p^*_{GWR}$ in many cases.

For GWR, it is important to note that the singularity of $\mathbf{X}'\mathbf{G}(s_i)\mathbf{X}$ changes depending on the spatial scale of the predictor variables. If $\mathbf{x}_k$ suggests small-scale spatial variations, it would have much variations within each kernel window. By contrast, if $\mathbf{x}_k$ suggests large-scale spatial variations, its variations within each kernel window can be small. In the extreme case, if $\mathbf{x}_k$ has uniform values across the window around site $s_i$, it is exactly collinear with the intercept term within the window (i.e., suppose that $\mathbf{x}_1$ represents an intercept, the entries of the $k$-th row and column of $\mathbf{X}'\mathbf{G}(s_i)\mathbf{X}$ take exactly the same values with the entries of the 1-st row and column. The resulting $\mathbf{X}'\mathbf{G}(s_i)\mathbf{X}$ becomes singular). The FB-GWR model, which calibrates the bandwidths implicitly considering the scale of each $\mathbf{x}_k$, is valuable not only to control the varying scales of the SVCs, but also to stabilize the SVC estimates (e.g., in the presence of collinearity).

Regarding ESF, forward eigenvector selection implicitly identifies the model that maximizes accuracy, where $p^*_{ESF} = tr[\mathbf{H}_{ESF}]$. Given the fact that the Moran eigenvectors describe coefficient patterns at different spatial scales, eigenvector selection identifies the scale of spatial variation in each SVC set. $p^*_{ESF}$ increases as the number of selected eigenvectors increases; it happens when SVCs have spatial variations in every scale.

Unlike all of the above models, the effective number of parameters for the RE-ESF model, $p^*_{RE-ESF} = tr[\mathbf{H}_{RE-ESF}]$, includes not just the scale parameters $\{\alpha_1,...\alpha_K\}$, but also the variance parameters $\{\sigma^2_{1,\gamma},...\sigma^2_{K,\gamma}\}$. Different from GWR, FB-GWR and ESF models whose $p^*$ always increase when SVCs have small-scale variations, the RE-ESF model is capable of stabilizing $p^*_{RE\text{-}ESF}$ even if the SVCs tend have small-scale spatial patterns (i.e., small $\alpha_1,...\alpha_K$) by decreasing the variance parameters.

## 3. Monte Carlo simulation 1: scale vs. model complexity

### 3.1. Outline

This section objectively evaluates model complexity with $p^*$ values, while varying the predictor variables and the scale parameters for the SVCs, and tests for which cases the SVC models are unstable (i.e., investigates model limitation (a), from above). For simplicity, we evaluate only cases where the spatial scale of variation for each set of SVCs is the same. In other words, regression relationships are set to vary from the small-scale to the large-scale, but always in the same fashion for each regression relationship in the model. Thus FB-GWR is not analyzed here, as it simply defaults to standard GWR in this instance.

For the Monte Carlo simulation, we assume a SVC model, $\mathbf{y} = \boldsymbol{\beta}_0 + \mathbf{x}_1 \circ \boldsymbol{\beta}_1 + \mathbf{x}_2 \circ \boldsymbol{\beta}_2 + \boldsymbol{\varepsilon}$, where the predictor variables are generated from:

$$\mathbf{x}_k = (1 - r_x)\boldsymbol{\varepsilon}_{x(ns)} + r_x \mathbf{C}(b_x)\boldsymbol{\varepsilon}_{x(s)}, \tag{14}$$

where $\boldsymbol{\varepsilon}_{x(ns)} \sim N(\mathbf{0}, \mathbf{I})$ and $\boldsymbol{\varepsilon}_{x(s)} \sim N(\mathbf{0}, \mathbf{I})$. Here $\mathbf{C}(b_k)$ is a matrix that row-standardizes a symmetric spatial proximity matrix whose $(i, j)$-th element equals $\exp(-d(s_i, s_j)/b_k)$, and where $d(s_i, s_j)$ is the Euclidean distance between sample sites $s_i$ and $s_j$. Spatial coordinates of the sample sites are also allowed to vary, and are generated from standard normal distributions. Thus, $\mathbf{C}(b_k)\boldsymbol{\varepsilon}_k$ is a spatial moving average process, and $r_x$ is the ratio of spatially dependent variation to total variation in $\mathbf{x}_k$. Therefore, $p^*$ values for GWR, GWRa, ESF, and RE-ESF are found while varying the parameters for the predictor variables (see Table 2) and those for the SVCs (see Table 3). In each case, the $p^*$ values of each model are evaluated 200 times. For ESF, $[\mathbf{x}_1 \circ \mathbf{E}_1 \quad ... \quad \mathbf{x}_K \circ \mathbf{E}_K]$, where $\mathbf{E}_k$ consists of eigenvectors corresponding to positive eigenvalues, are candidates for the variable selection step. These eigenvectors are also used in RE-ESF (i.e., $\mathbf{E} = \mathbf{E}_k$). The ratio of the selected eigenvectors in ESF are given as described in Table 3.

Note that this section does not estimate SVCs through model fitting, but calculates $p^*$ by simply substituting know parameters into $p^*=tr[\mathbf{H}]$ (see, Table 1). For SVC estimation accuracy, see Section 4.

**[Table 2 around here]**

**[Table 3 around here]**

### 3.2. Results

Figure 1 plots the mean estimated $p^*$ values for the SVC models arising from this first Monte Carlo simulation experiment. Here the mean $p^*_{GWR}$ results suggest that GWR (with a fixed bandwidth) is unstable when the SVCs have small-scale spatial variation ($b = 0.2$). In other words, GWR could be stable unless the fixed bandwidth is inappropriately small (i.e., $b \neq 0.2$).

Unlike the mean $p^*_{GWR}$ results, the mean $p^*_{GWRa}$ results for GWRa (with its adaptive bandwidth) are always relatively small across all values of $b$. At least from this result, GWRa seems relatively stable compared to GWR. This is not surprising, given that numerous empirical studies have suggested as much (e.g., Harris et al., 2010a). Only for highly regular sample configurations is fixed bandwidth GWR usually recommended.

The drawback to GWRa, however, is that it implies non-stationary relationships are operating within their own local region of dependence, whilst fixed bandwidth GWR, ensures these regions are the same size everywhere, and thus provides more generalized interpretations of the geographical process under study. For example, reporting that the nature of the relationship between crime and unemployment depends only on incident characteristics within a 2km radius of the crime scene is intuitively more informative than reporting that this relationship depends only on the characteristics of the nearest 30 incidents of the crime scene.

**[Figure 1 around here]**

In Figure 1, the mean $p^*_{RE\text{-}ESF}$ results are evaluated for cases with $\sigma_k =0.1$ and $\sigma_k =1.0$, respectively. On the one hand, when $\sigma_k =1.0$, which implies weaker shrinkage, $p^*_{RE\text{-}ESF}$ takes large values. On the other hand, $p^*_{RE\text{-}ESF}$ values are small across cases when stronger shrinkage is imposed by $\sigma_k =0.1$. Thus, the RE-ESF estimates are relatively stable even when the SVCs have local variation, but with a proviso that the $\sigma_k$ parameter is estimated appropriately. Furthermore, the mean $p^*_{ESF}$ values, which equal the number of selected predictor variables, become {26.8, 50.7, 75.5, 98.3} in cases where the ratio of selected eigenvectors equals {0.2, 0.4, 0.6, 0.8}, respectively. Considering the usefulness of the shrinkage parameter, $\sigma_k$, in RE-ESF, regularized ESF (e.g., Seya et al., 2011) might be useful to reduce $p^*_{ESF}$.

In summary, fixed bandwidth GWR can be very unstable when the bandwidth of local parameter estimation is inappropriately small, and ESF tends to be unstable as the number of selected eigenvectors increase. Conversely, GWRa is stable across both small- and large-scale SVC processes, and RE-ESF is similarly stable, provided that $\sigma_k$ is estimated appropriately.

Also observable from Figure 1 is that predictor variable, $\mathbf{x}_k$, with small-scale spatial variations universally make all SVC models unstable. In contrast, in a context of global spatial regression (e.g., spatial error model; e.g., LeSage and Pace, 2009), Paciorek (2010) analytically showed that the coefficient estimates tend to be unstable if the spatial scales of the predictor variables are larger than the scale of the residual spatial process.[3] Thus, both small-scale $\mathbf{x}_k$ and large-scale $\mathbf{x}_k$ influence the reliability of the SVC estimates for different reasons. The next section investigates the influence of all these instabilities on the accuracy of the SVC estimates themselves, and tries to determine whether small-scale $\mathbf{x}_k$ or large-scale $\mathbf{x}_k$ is more harmful in SVC estimation.

## 4. Monte Carlo simulation 2: scale vs. SVC estimation accuracy

### 4.1. Outline

This section compares all six study SVC models (GWR, GWRa, FB-GWR, FB-GWRa, ESF and RE-ESF) though another Monte Carlo simulation experiment, where we now assess the accuracy of the estimated SVCs in relation to the (known) simulated SVCs. The synthetic data are generated from the following SVCs model:

$$\mathbf{y} = \boldsymbol{\beta}_0 + \mathbf{x}_1 \circ \boldsymbol{\beta}_1 + \mathbf{x}_2 \circ \boldsymbol{\beta}_2 + \boldsymbol{\varepsilon}, \qquad \boldsymbol{\varepsilon} \sim N(\mathbf{0},\ 2^2 \mathbf{I}), \tag{15}$$

---

[3] Because SVCs are assumed known in section 4, the instability in their estimation does not appear in here.

$$\boldsymbol{\beta}_0 = \mathbf{1} + \mathbf{C}(b_0)\boldsymbol{\varepsilon}_0, \quad \boldsymbol{\beta}_1 = (-2)\mathbf{1} + 3\mathbf{C}(b_1)\boldsymbol{\varepsilon}_1, \quad \boldsymbol{\beta}_2 = (0.5)\mathbf{1} + \mathbf{C}(b_2)\boldsymbol{\varepsilon}_2,$$

where $\boldsymbol{\varepsilon}_k \sim N(\mathbf{0}, \mathbf{I})$. The spatial variation in $\boldsymbol{\beta}_1$ is three times stronger than the spatial variation in $\boldsymbol{\beta}_0$ and $\boldsymbol{\beta}_2$. We refer to $\boldsymbol{\beta}_1$ as a significant SVC process, whilst $\boldsymbol{\beta}_0$ and $\boldsymbol{\beta}_2$ are considered to be insignificant SVC processes. Thus this simulation experiment specifically investigates model limitation (b), from above. Following the previous section, the predictor variables are generated from $\mathbf{x}_k = (1 - r_x)\boldsymbol{\varepsilon}_{x(ns)} + r_x\mathbf{C}(b_x)\boldsymbol{\varepsilon}_{x(s)}$. Parameters are estimated 200 times while varying parameter values, as summarized in Table 4.

[Table 4 around here]

4.2. Results

The accuracy of each model's SVC estimates is evaluated using root mean squared error (RMSE), mean absolute error (MAE), and bias diagnostics. The results and explanations of the MAE and bias diagnostics are given in the Appendix, whilst only the RMSE results are presented here. The RMSE for estimated $\widehat{\boldsymbol{\beta}}_\mathbf{k}$ is given by the mean of RMSE$[\hat{\beta}_k(s_i)]$, which is formulated as follows:

$$RMSE[\hat{\beta}_k(s_i)] = \sqrt{\frac{1}{200}\sum_{iter=1}^{200}(\beta_k(s_i) - \hat{\beta}_k(s_i))^2}. \tag{16}$$

where $\beta_k(s_i)$ is the true SVC value generated from Eq. (15). To visualize the simulation results effectively, we use 2-dimentional plots as presented in Figures 2 to 6. Here the horizontal axis always denotes the RMSEs for RE-ESF, whose SVC estimation accuracy has been shown to be relatively good across all cases in the 'companion study' of Murakami et al. (2017), and the vertical axis denotes the RMSEs for one of the models GWR, ESF, FB-GWR, GWRa, and FB-GWRa.

Figure 2 compares (fixed bandwidth) GWR with RE-ESF for SVC accuracy via RMSE. Here, GWR provides more accurate SVCs than RE-ESF if the plot outputs are concentrated in the bottom right triangle of each panel, while the estimated SVCs from RE-ESF are more accurate if the plot outputs are in the top left triangle. Results clearly demonstrates that the RMSEs from GWR are generally greater than those from RE-ESF, and thus RE-ESF tends to be more accurate. This tendency is most conspicuous when the significant SVC ($\beta_1$) has the small-scale variables and $\mathbf{x}_s$ has strong large-scale variations, verifying that different scales of relationship non-stationarity need to be accounted for in SVC models (which RE-ESF does, but GWR does not). This tendency is also substantial for the largest sample size, when $N = 400$. By contrast, GWR can perform equally as well, or better than, RE-ESF when $N = 50$. This is interesting, and may suggest that the smaller the sample size, the more difficult it is to detect relationships varying locally and across different spatial scales. Furthermore, Páez et al. (2011) recommended using GWR only for large samples ($N > 160$), but these results suggest some value in GWR for small samples. Figure 3 compares the coefficient RMSE results for ESF with RE-ESF, where it is clear that ESF provides poorer levels of SVC estimation accuracy than RE-ESF, across all nine scenarios. As with GWR (and as would be expected), ESF provides relatively inaccurate SVCs in cases with small-scale variations in the significant SVC ($\boldsymbol{\beta}_1$) and strong large-scale variations in $\mathbf{x}_s$. Although not shown graphically, by comparing Figure 2 with Figure 3, it is strongly suspected that GWR tends to perform better than ESF.

**[Figure 2 around here]**

**[Figure 3 around here]**

Figure 4 compares FB-GWR and RE-ESF, where, interestingly, unlike GWR, no singular estimates appear from FB-GWR. Thus, GWR with multiple bandwidths (in this FB-GWR form) appears to stabilize SVC estimates, and tentatively may provide a useful alternative to a regularized GWR model in addressing local collinearity issues. As most plots follow the 45º line in Figure 4, FB-GWR provides SVC estimates that tend to be just as accurate as those from RE-ESF. Moreover, FB-GWR SVC estimates are more accurate than RE-ESF when $N = 50$. This is because RE-ESF is a likelihood approach relying on the law of large numbers. Conversely, the SVC estimates for FB-GWR tend to be marginally less accurate than those from RE-ESF when $\mathbf{x}_k$ have strong large-scale spatial variations, and the significant SVC ($\boldsymbol{\beta}_1$) has small-scale variations (but for $N = 150$ and for $N = 400$, only).

**[Figure 4 around here]**

Figure 5 compares GWRa with RE-ESF for SVC accuracy. Somewhat surprisingly, GWRa does not suffer from any singular fit, and RMSE values are greatly reduced compared to the fixed bandwidth GWR results in Figure 2. The use of an adaptive bandwidth appears to be a simple and efficient solution to stabilize GWR modeling, although, in this case, stability may relate more to the effects of sample configurations than to other influences. Conversely, GWRa provides much poorer levels of SVC accuracy (than GWR and RE-ESF) for the significant small-scale SVC, $\boldsymbol{\beta}_1$. Furthermore, the GWRa estimates for insignificant SVCs, $\boldsymbol{\beta}_0$ and $\boldsymbol{\beta}_2$, tend to be more accurate than that found for GWR (see Figure 2), whilst the GWRa estimates provide broadly similar levels of accuracy to that found for RE-ESF for such cases. Figure 6 compares FB-GWRa with RE-ESF for SVC accuracy. Here FB-GWRa appears to have similar coefficient accuracy tendencies to those found with both FB-GWR (Figure 4) and GWRa (Figure 5), in relation to RE-ESF. As would be expected, FB-GWRa is more accurate than GWRa for the significant small-scale SVC, $\beta_1$, where the FB-GWRa results are more compatible with those from RE-ESF. Overall, FB-GWRa is found to estimate both weak and strong SVC processes relatively accurately.

**[Figure 5 around here]**

**[Figure 6 around here]**

Finally, Table 5 compares average computational times for all six SVC models for the three sample sizes of $N = 50$, 150, and 400. As expected, GWR and GWRa run the fastest, as they are relatively simple. Considering the RMSE accuracy results, above, it is recommended that GWRa would often be a sensible and pragmatic choice for very large datasets. By contrast, FB-GWR and FB-GWRa are relatively slow due to their usage of the back-fitting algorithm in their calibration. Acceleration of these multi-scale GWR models would be an important research topic in the future, although some work in this area is currently in progress (Lu et al. in review). The ESF model is also slow because it requires stepwise eigenvector selection. By contrast, RE-ESF is as fast as GWR and GWRa, despite the fact that it estimates each spatial scale of each set of SVCs (i.e., RE-ESF is multi-scale). This is because the computational complexity for optimizing the scale parameters is only $O(L^3K^3)$, which is independent of sample size. Note also that the cost for eigen-decomposition for RE-ESF, which is severe when $N$ is large, can be lightened dramatically by an approximation proposed by Griffith (2000), which is for regular lattice data, or Murakami and Griffith (2017). Thus, RE-ESF is also recommended for very large datasets, and should be preferred to GWRa when relationships are not only expected to vary locally, but also across different spatial scales.

**[Table 5 around here]**

## 5. Concluding remarks

The study summarized in this paper investigated the influence of scale on SVC modeling, where relationships between the response and predictor not only operate locally, but also at varying spatial scales. Results from simulation experiments suggest that standard GWR provides poor SVC estimates, when some SVCs vary at a small-scale whilst others vary at a large-scale. By contrast, a multi-scale GWR model in FB-GWR provides SVC estimates that are relatively accurate for such processes. Interestingly, differences in SVC estimation accuracy, and model stability, also depend on whether fixed distance or adaptive distance kernel bandwidths are specified for GWR or for FB-GWR, where adaptive ones should in general be preferred.

GWR and FB-GWR are examples of *local* approaches to SVC modelling, whilst ESF and RE-ESF are both *global* approaches. Here RE-ESF is a regularized ESF model that is designed to capture scale dependencies in local relationships, just as FB-GWR is. The RE-ESF is shown to more accurately estimate such multi-scale SVC processes in comparison to not only ESF, but also to GWR. RE-ESF is also shown to be a more stable model than ESF or GWR. Both FB-GWR and RE-ESF are found to provide the most accurate estimates of the SVC processes generated in the simulation experiment, but where RE-ESF is shown to be the most computationally efficient, and thus more suitable for very large datasets. Overall, the results strongly indicate that future SVC studies need to pay more attention to issues of spatial scale, and investigate with a FB-GWR or RE-ESF model, especially considering it is entirely unrealistic for each set of SVCs to operate at the same spatial scale (as naively assumed in standard GWR or ESF models).

Still, there are some remaining issues, where future work on the analytic properties of spatial scale and SVC estimates could follow that of Paciorek (2010), where only the effects on stationary regression coefficients were investigated. Such studies would help in understanding the scale problem more deeply, and possibly enable the establishment of a local/global indicator of scale dependence for SVCs. Extensions should also consider: (i) non-Gaussian data modeling (Atkinson et al., 2003; Griffith, 2002; 2004; Nakaya et al., 2005), (ii) spatiotemporal modeling (Huang et al., 2010; Griffith, 2012; Fotheringham et al., 2015), (iii) spatial prediction (Harris et al., 2010b; 2011; Griffith, 2013), (iv) spatial interaction modeling (Nakaya, 2001; Kordi and Fotheringham 2016; Griffith et al., 2017), and (v) the mitigation of the modifiable areal unit problem (Fotheringham et al., 2002; Murakami and Tsutsumi, 2015).

## Acknowledgements

This study was jointly funded by the JSPS KAKENHI Grant Numbers 17K12974 and 17K14738; projects from the National Natural Science Foundation of China [NSFC: 41401455 and U1533102]. Work was also supported by a UK Biotechnology and Biological Sciences Research Council grant (BBSRC BB/J004308/1).

- Wheeler, D. C., & Calder, C. A. (2007). An assessment of coefficient accuracy in linear regression models with spatially varying coefficients. *Journal of Geographical Systems*, 9 (2), 145-166.

- Wheeler, D. C., & Waller, L. A. (2009). Comparing spatially varying coefficient models: a case study examining violent crime rates and their relationships to alcohol outlets and illegal drug arrests. *Journal of Geographical Systems*, 11 (1), 1-22.

- Wheeler, D., & Tiefelsdorf, M. (2005). Multicollinearity and correlation among local regression coefficients in geographically weighted regression. *Journal of Geographical Systems*, 7 (2), 161-187.

- Yang W, Fotheringham AS, Harris P (2011) Model selection in GWR: the development of a flexible bandwidth GWR. Geocomputation 2011, London, UK

- Yang W, Fotheringham AS, Harris P (2012) An extension of Geographically Weighted Regression with Flexible Bandwidths. GISRUK 2012, Lancaster, UK

- Yang, W. (2014). *An extension of geographically weighted regression with flexible bandwidths*. Doctoral dissertation, University of St Andrews.42

Table 1: The hat matrix, **H**, of the study SVC models, where $p^* = \text{tr}[\mathbf{H}]$.

| Model | Hat matrix, **H** | Parameters in **H** |
|---|---|---|
| GWR | $\mathbf{H}_{GWR}$ = a matrix with its $i$-th row being: $\mathbf{x}(s_i)'[\mathbf{X}'\mathbf{G}(s_i)\mathbf{X}]^{-1}\mathbf{X}'\mathbf{G}(s_i)$ | $b$ |
| FB-GWR | Single hat matrix is not available (Fotheringham et al., 2017). The following hat matrix for $k$-th SVC appears in each iteration of the backfitting: $\mathbf{x}(s_i)'[\mathbf{x}_k'\mathbf{G}_k(s_i)\mathbf{x}_k]^{-1}\mathbf{x}_k'\mathbf{G}_k(s_i)$, where $\mathbf{G}_k(s_i)$ equals $\mathbf{G}(s_i)$ whose $b$ is replaced with $b_k$. | $b_1,\ldots b_K$ |
| ESF | $\mathbf{H}_{ESF} = [\mathbf{X} \quad \tilde{\mathbf{E}}_{ESF}] \begin{bmatrix} \mathbf{X}'\mathbf{X} & \mathbf{X}'\tilde{\mathbf{E}}_{ESF} \\ \tilde{\mathbf{E}}'_{ESF}\mathbf{X} & \tilde{\mathbf{E}}'_{ESF}\tilde{\mathbf{E}}_{ESF} \end{bmatrix}^{-1} \begin{bmatrix} \mathbf{X}' \\ \tilde{\mathbf{E}}'_{ESF} \end{bmatrix}$, where $\tilde{\mathbf{E}}_{ESF} = [\mathbf{x}_1 \circ \mathbf{E}_1 \ \ldots \ \mathbf{x}_K \circ \mathbf{E}_K]$ | Selection of eigenvectors |
| RE-ESF | $\mathbf{H}_{RE-ESF} = [\mathbf{X} \quad \tilde{\mathbf{E}}\tilde{\mathbf{\Lambda}}(\mathbf{\theta})] \begin{bmatrix} \mathbf{X}'\mathbf{X} & \mathbf{X}'\tilde{\mathbf{E}}\tilde{\mathbf{\Lambda}}(\mathbf{\theta}) \\ \tilde{\mathbf{\Lambda}}(\mathbf{\theta})\tilde{\mathbf{E}}'\mathbf{X} & \tilde{\mathbf{\Lambda}}(\mathbf{\theta})\tilde{\mathbf{E}}'\tilde{\mathbf{E}}\tilde{\mathbf{\Lambda}}(\mathbf{\theta}) + \mathbf{I}_{KL} \end{bmatrix}^{-1} \begin{bmatrix} \mathbf{X}' \\ \tilde{\mathbf{\Lambda}}(\mathbf{\theta})\tilde{\mathbf{E}}' \end{bmatrix}$ | $\mathbf{\theta} \in \{\alpha_1,\ldots\alpha_K, \sigma^2_{1,\gamma},\ldots\sigma^2_{K,\gamma}\}$ |

Table 2: Parameter settings for predictor variables: $\mathbf{x}_k$

| Parameter | Notation | Case |
|---|---|---|
| Sample size | $N$ | 400 |
| Bandwidth | $b_x$ | {0.0, 0.2, 0.6, 1.0} |
| Ratio of spatial variation | $r_x$ | {0.2, 0.6, 1.0, 2.0} |

Table 3: Parameter settings for SVCs: $\mathbf{\beta}_k$

| Model | Parameter | Notation | Case |
|---|---|---|---|
| GWR | Bandwidth | $b$ | {0.2, 0.6, 1.0, 2.0} |
| GWRa | Adaptive bandwidth | $b^{ad}$ | {0.1, 0.3, 0.5, 1.0} |
| ESF | Ratio of predictor variables being selected | $q$ | {0.2, 0.4, 0.6, 0.8} |
| RE-ESF | Scale | $\alpha_k$ | {0.2, 0.6, 1.0, 2.0} |
| | Variance | $\sigma_k$ | {0.1, 1.0} |

Table 4: Parameter settings in 144 ($3 \times 4 \times 3 \times 4$) cases.

| Parameter | Notation | Case |
|---|---|---|
| Sample size | $N$ | {50, 150, 400} |
| Bandwidth for {$\beta_0$, $\beta_1$, $\beta_2$} | ($b_0$, $b_1$, $b_2$) | {(0.2, 0.2, 0.2), (1.0, 0.2, 1.0), (0.2, 1.0, 0.2), (1.0, 1.0, 1.0)} |
| Bandwidth for $\mathbf{x}_k$ | $b_x$ | {0.2, 0.6, 1.0} |
| Ratio of spatial variation in $\mathbf{x}_k$ | $r_x$ | {0.0, 0.4, 0.8, 1.0} |

Table 5: Average computational time in seconds.

| $N$ | GWR | GWRa | FB-GWR | FB-GWRa | ESF | RE-ESF |
|---|---|---|---|---|---|---|
| 50 | 0.13 | 0.18 | 1.50 | 10.31 | 1.49 | 0.29 |
| 150 | 0.54 | 0.72 | 12.02 | 12.96 | 10.52 | 0.77 |
| 400 | 2.44 | 2.63 | 93.52 | 65.41 | 72.56 | 3.38 |

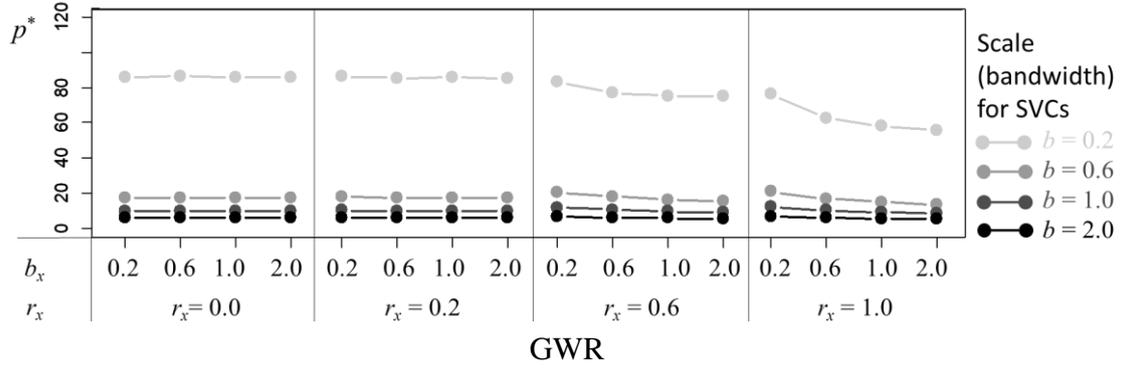
GWR

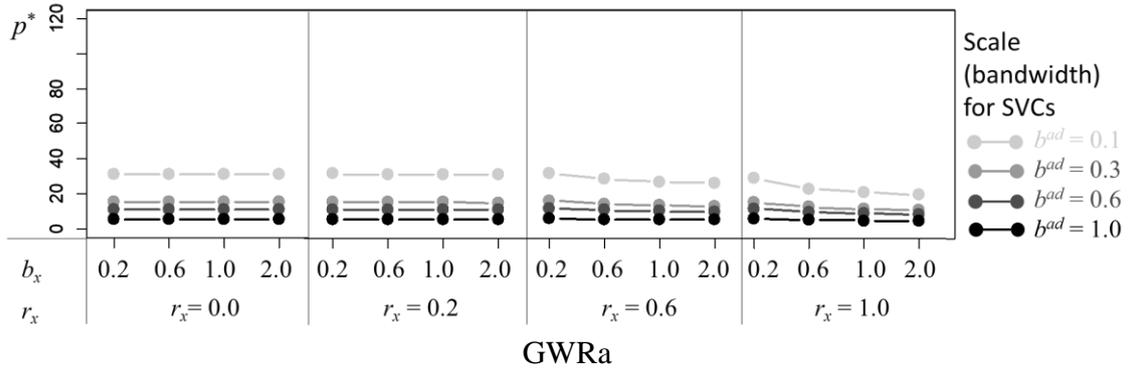
GWRa

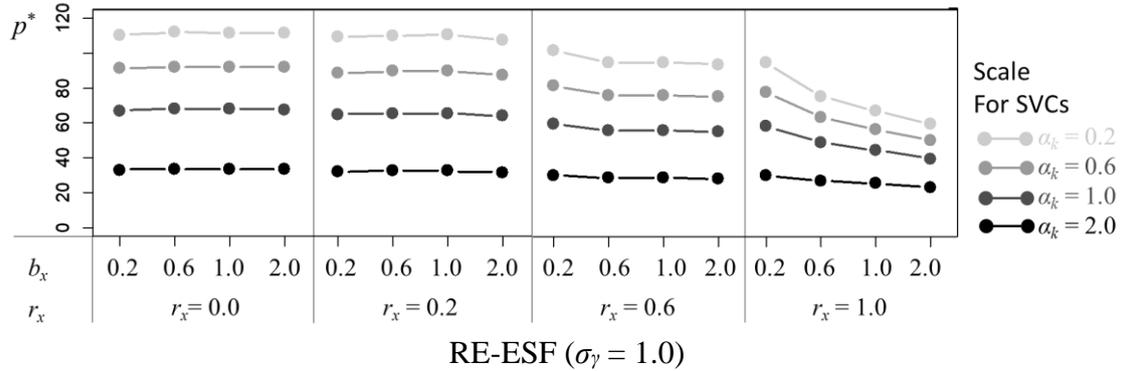
RE-ESF ($\sigma_\gamma = 1.0$)

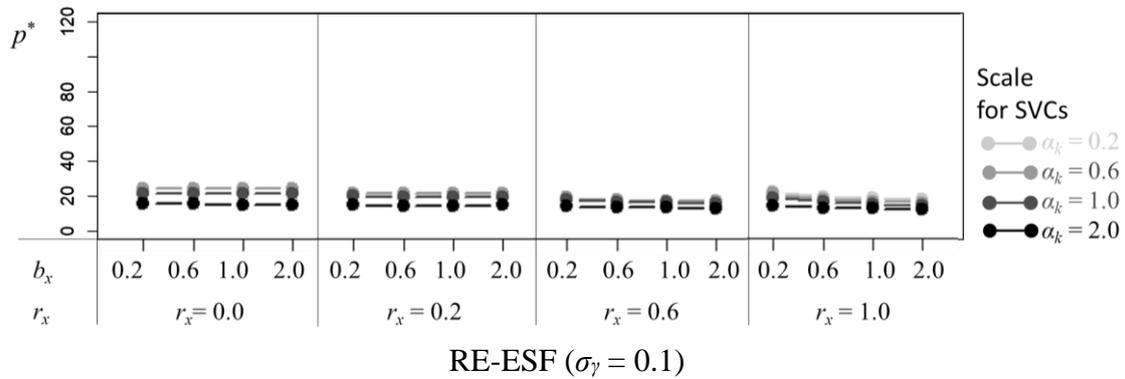
RE-ESF ($\sigma_\gamma = 0.1$)

Figure 1: Mean effective number of parameters, $p^*$, with respect to the scale of the SVCs. In each panel, lighter lines represent $p^*$s evaluated with more locally-tending SVCs, whilst darker lines represent $p^*$s evaluated with more globally-tending SVCs.

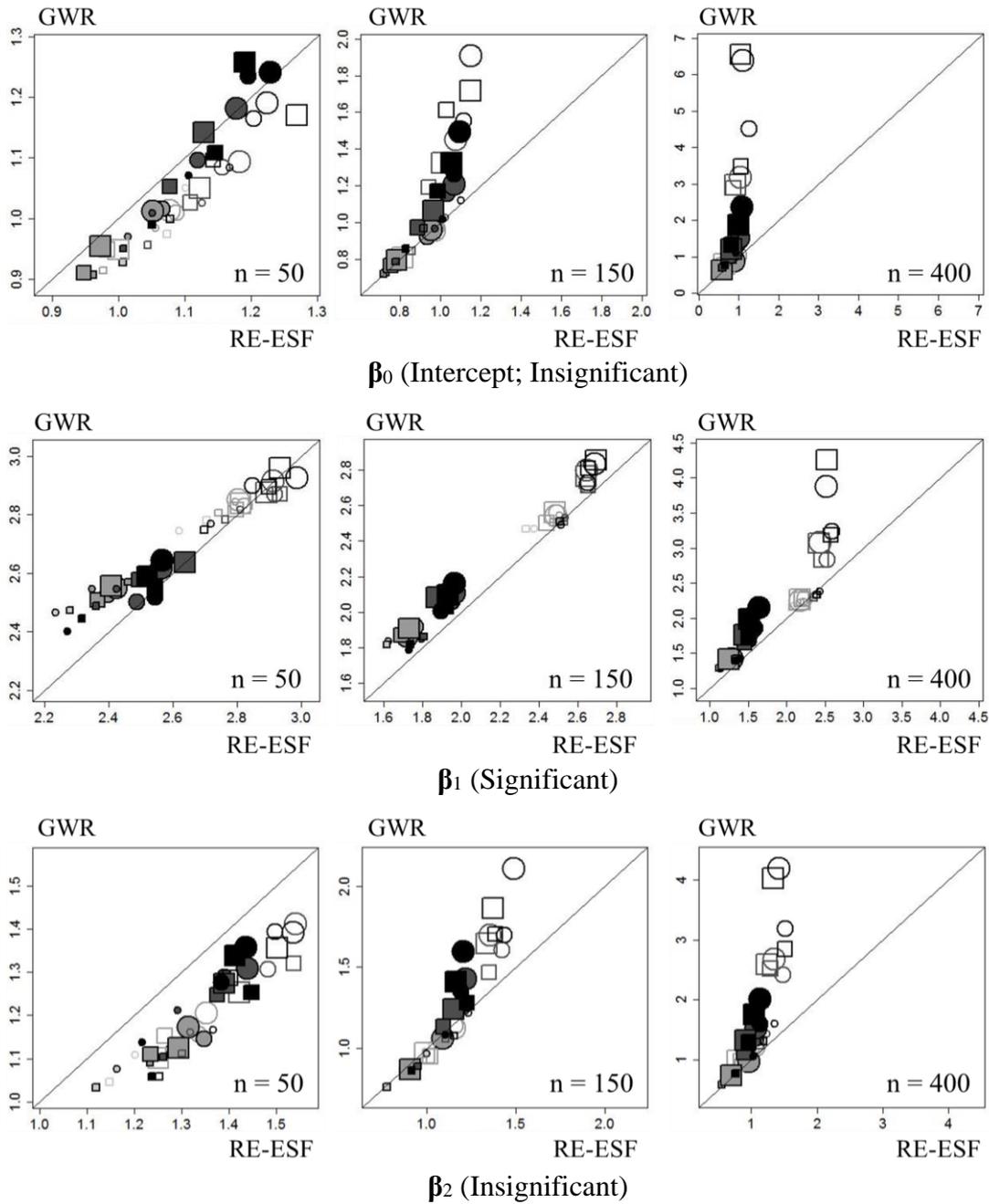

Figure 2: RMSE: RE-ESF (*x*-axis) vs GWR (*y*-axis). "Large-scale" means the large-scale ($r$). The lighter end of the "Significant" line means a small variance of the spatially dependent component ($s_x$), while the darker end means a large variance.

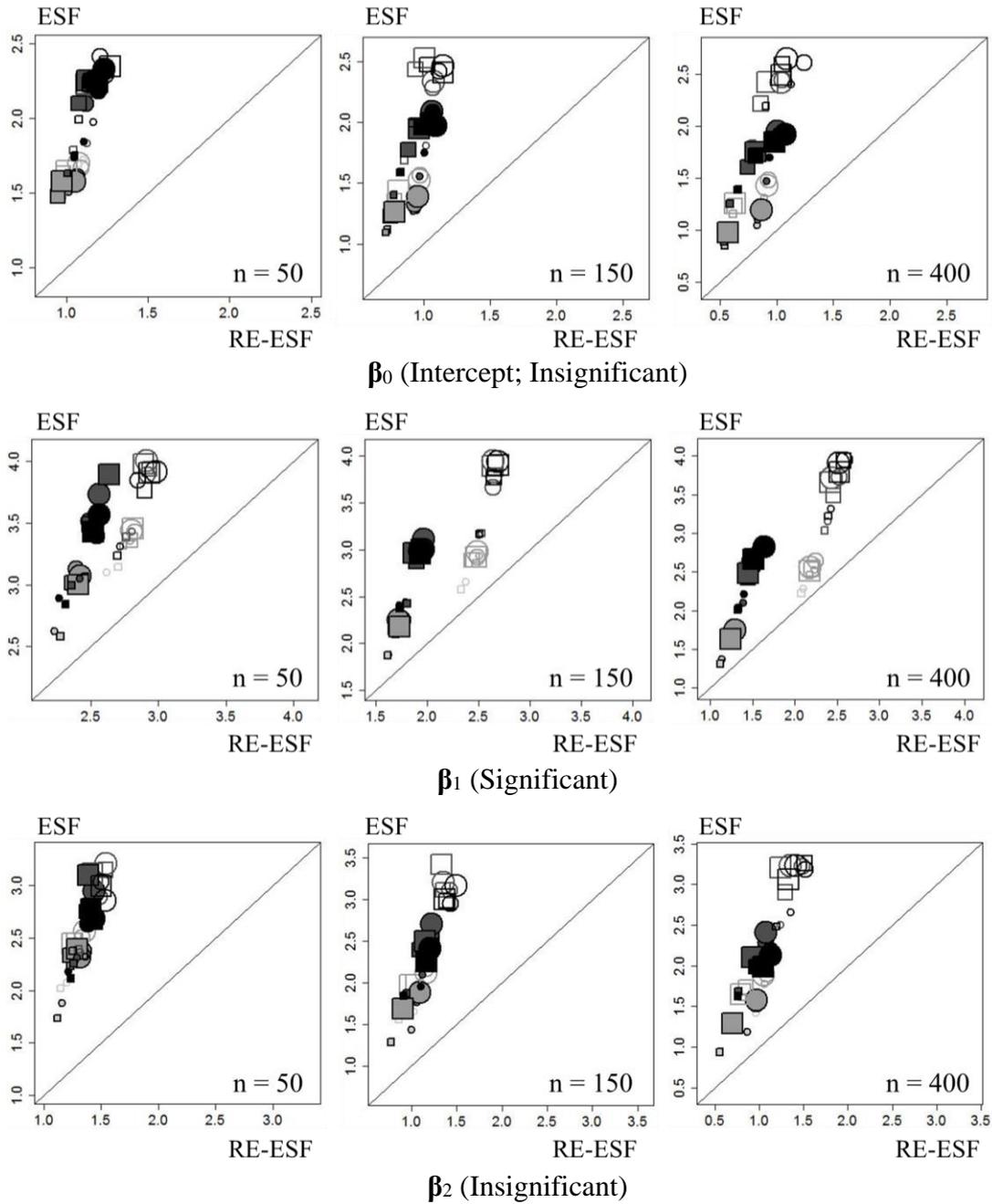

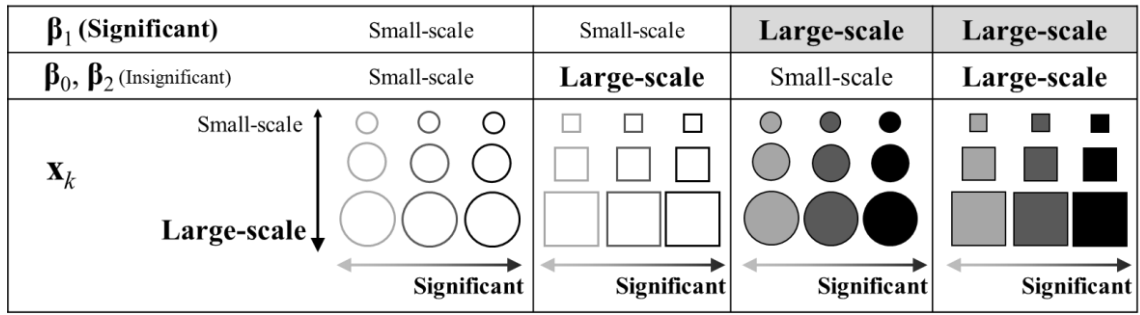

Figure 3: RMSE: RE-ESF (*x*-axis) vs ESF (*y*-axis). "Large-scale" means the large-scale (*r*). The lighter end of the "Significant" line means a small variance of the spatially dependent component ($s_x$), while the darker end means a large variance.

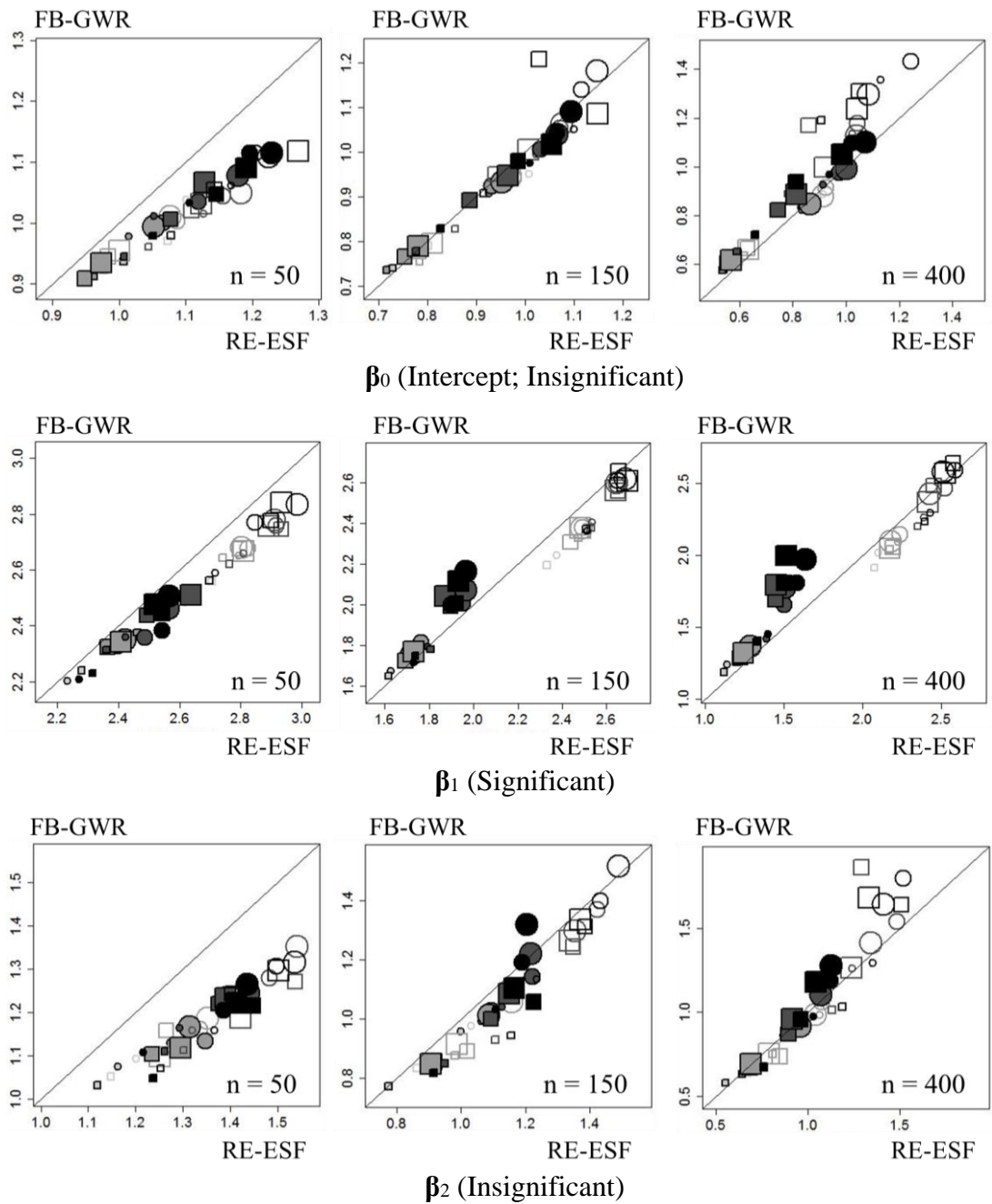

Figure 4: RMSE: RE-ESF (*x*-axis) vs FB-GWR (*y*-axis). "Large-scale" means the large-scale ($r$). The lighter end of the "Significant" line means a small variance of the spatially dependent component ($s_x$), while the darker end means a large variance.

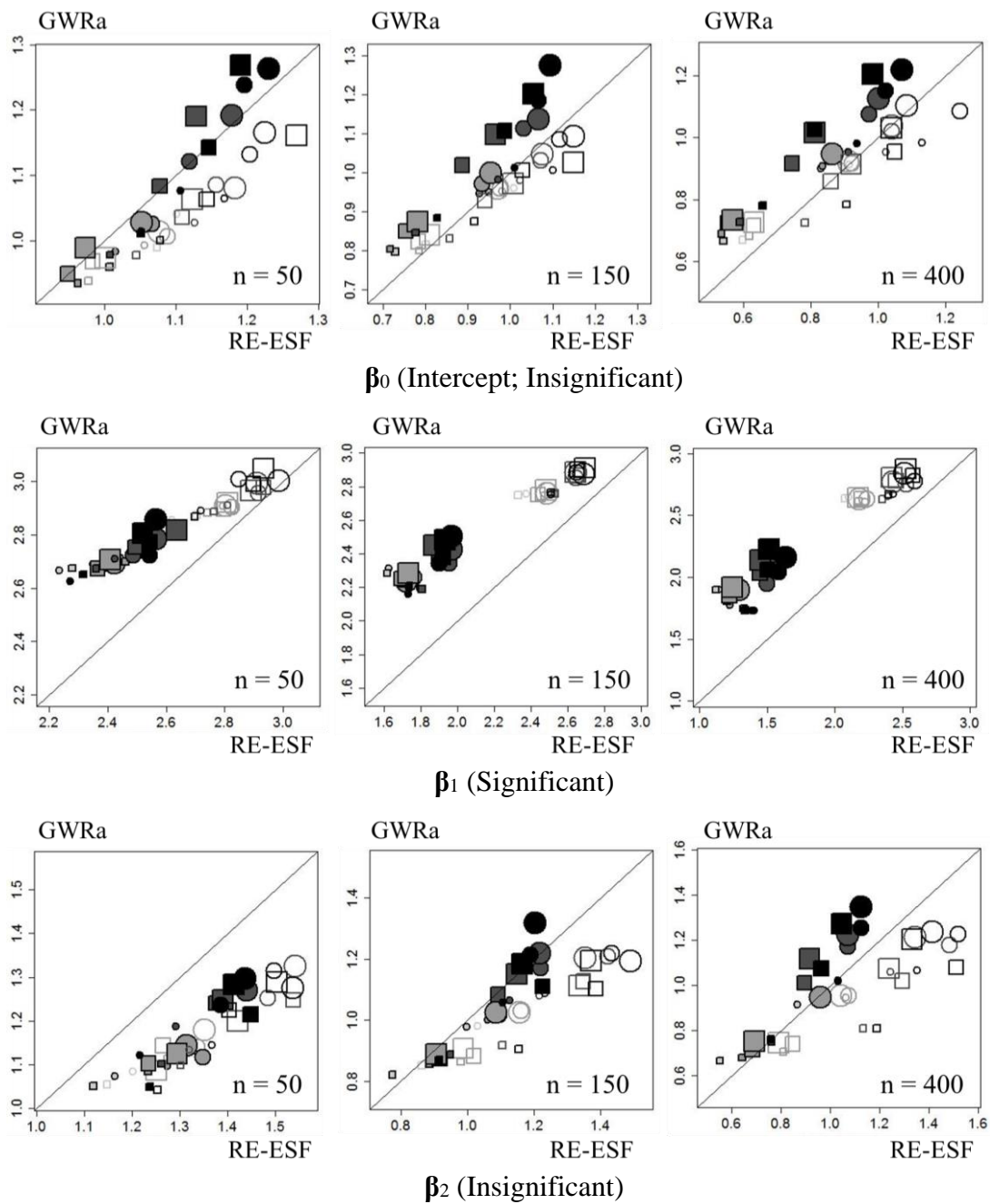

Figure 5: RMSE: RE-ESF (*x*-axis) vs GWRa (*y*-axis). "Large-scale" means the large-scale ($r$). The lighter end of the "Significant" line means a small variance of the spatially dependent component ($s_x$), while the darker end means a large variance.

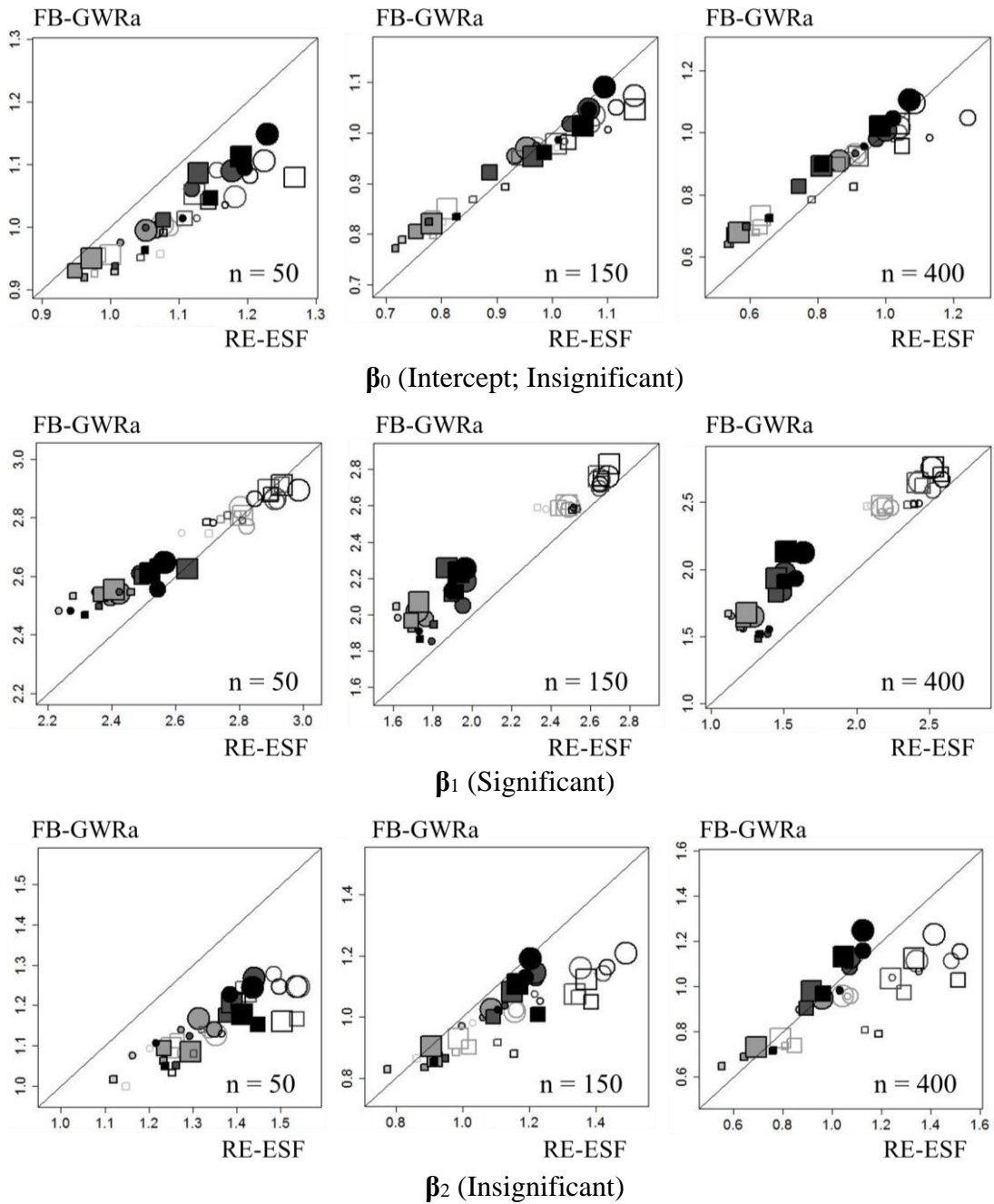

Figure 6: RMSE: RE-ESF ($x$-axis) vs FB-GWRa ($y$-axis). "Large-scale" means the large-scale ($r$). The lighter end of the "Significant" line means a small variance of the spatially dependent component ($s_x$), while the darker end means a large variance.